\documentclass[12pt]{article}
\pdfoutput=1
\usepackage{amssymb}
\usepackage{amsmath}
\usepackage{graphicx}
\usepackage{hyperref}
\usepackage{color}
\usepackage{fullpage}

\numberwithin{equation}{section}

\newcommand{\eq}{\begin{equation}}
\newcommand{\eqx}{\end{equation}}
\newcommand{\eqs}{\begin{equation*}}
\newcommand{\eqsx}{\end{equation*}}
\newcommand{\eqn}{\begin{eqnarray}}
\newcommand{\eqnx}{\end{eqnarray}}
\newcommand{\eqns}{\begin{eqnarray*}}
\newcommand{\eqnsx}{\end{eqnarray*}}

\newcommand{\f}[2]{\frac{#1}{#2}}

\newcommand{\ket}[1]{\left|{#1}\right\rangle}

\newcommand{\lm}{\lambda}

\renewcommand{\th}{\theta}
\newcommand{\sg}{\sigma}

\newcommand{\dl}{\delta}

\newcommand{\Gm}{\Gamma}

\newcommand{\om}{\omega}

\newcommand{\gm}{\gamma}

\newcommand{\eps}{\varepsilon}
\newcommand{\qqqq}{\quad\quad\quad\quad}
\newcommand{\qq}{\quad\quad}

\newcommand{\nn}{{\cal N}}

\DeclareMathOperator{\dn}{dn}
\DeclareMathOperator{\sn}{sn}
\DeclareMathOperator{\cn}{cn}
\DeclareMathOperator{\am}{am}

\newcommand{\Nfin}{\mathbf{N}}

\newcommand{\Gmel}{\Gamma_{\!ell}}

\newcommand{\ntl}{\tilde{n}}
\newcommand{\Gt}{\tilde{\Gamma}_{\f{ML}{2\pi}}}

\newcommand{\qth}[1]{\th_0 \left( #1 \right)}

\title{The kinematical $AdS_5 \times S^5$ Neumann coefficient}

\author{Zoltan Bajnok$^{a}$\thanks{e-mail: {\tt bajnok.zoltan@wigner.mta.hu}},\ \  
Romuald A. Janik$^{b}$\thanks{e-mail: {\tt romuald@th.if.uj.edu.pl}} \\ \\ 
\small 
${}^a$ MTA Lend\"ulet Holographic QFT Group\\\small
Wigner Research Centre\\\small
H-1525 Budapest 114, P.O.B. 49, Hungary\\\small
${}^b$ Institute of Physics\\\small
Jagiellonian University\\\small
ul. {\L}ojasiewicza 11, 30-348 Krak{\'o}w, Poland}

\date{}

\begin{document}

\maketitle

\begin{abstract}
For the case of two particles a solution of the string field theory vertex axioms
can be factorized into a standard form factor and a kinematical piece which includes
the dependence on the size of the third string.
In this paper we construct an exact solution of the kinematical axioms for $AdS_5 \times S^5$ which
includes all order wrapping corrections w.r.t. the size of the third string.
This solution is expressed in terms of elliptic Gamma functions and ordinary
elliptic functions.
The solution is valid at any coupling and we analyze its weak coupling, 
pp-wave and large $L$ limit.
\end{abstract}

\vfill

\pagebreak

\section{Introduction}

Recently there has been significant progress in our understanding of string interactions for string theories in
curved backgrounds which exhibit integrability. In our previous paper \cite{SFTUS} we formulated a set of functional equations
for the (light-cone) String Field Theory (SFT) three-string vertex for the case when the worldsheet theory is
integrable. The axioms \emph{per-se} apply to the case when two of the strings are large (more precisely they are decompactified)
while the third string can be of an arbitrary finite size $L$. The axioms depend in a nontrivial way on the size $L$.
The decompactification limit corresponds to cutting the string pants diagram (see Fig.~\ref{f.sftpants}) along one edge.
Since the third string has a finite size, the decompactification limit includes arbitrary number of wrapping corrections
w.r.t. $L$. This can be explicitly seen in the case of the pp-wave background geometry where we have at our disposal
an exact explicit solution for any value of $L$. Unfortunately we do not have, for the moment, a solution in the most interesting case 
of the $AdS_5 \times S^5$ geometry. This paper is a step in that direction.

In \cite{HEXAGON} a different approach was developed explicitly geared towards the computation of OPE coefficients in $\nn=4$ SYM.
Here the string vertex was cut along \emph{three} edges into two hexagons. This corresponds to the decompactification limit of
all three strings. In this context, functional equations for the hexagon in $AdS_5 \times S^5$ have been solved exactly.
The passage to finite volume incorporating wrapping effects involves, however, an iterative prescription for gluing
the hexagons together through integrating over an arbitrary number of particles on the edges being glued.
Thus wrapping effects are build on iteratively. Recently there appeared some further nontrivial checks of this proposal
\cite{SFONDRINITHREELOOPS,BKVTHREELOOPS} and it was even related \cite{Jiang:2015bvm} in the HHL ($L=0$) case to diagonal 
finite volume form factors. This is the structure which was conjectured in \cite{FFUS} and checked at weak coupling in 
\cite{Hollo:2015cda}.

In contrast, the finite $L$ solution of the SFT vertex axioms should at once resum an infinite set of wrapping corrections
and thus should provide some helpful information for the hexagon gluing procedure. 

In this paper we would like to find the simplest possible solutions of the SFT vertex axioms concentrating on
exactly treating the $L$ dependence. Of course any solution is given up to some analogs of CDD factors which
\emph{a-priori} can also be $L$ dependent (although the equations that they satisfy do not contain $L$). So
what we are aiming at is providing a `minimal' $L$ dependent solution. It will then remain an important
problem whether this solution is physical or whether it has to be suplemented by some additional CDD-like factors.
A similar question will arise for solutions for relativistic interacting integrable QFT's (e.g. sinh-Gordon
or the O(N) model on the decompactified pants diagram), which we will briefly also mention.
It would be very interesting to cross-check these simplest relativistic solutions in some other way and to understand
whether in that case any additional CDD-like factors are in fact necessary. This would
be important for our understanding of the required analytical structure.
Perhaps some integrable lattice realizations
of these integrable relativistic QFT's might shed light on these issues.

The plan of this paper is as follows. In section 2 we will briefly review the String Field Theory vertex axioms
proposed in \cite{SFTUS} and concentrate on the case of two particles relevant for the present paper.
Then we will review the structure of the pp-wave Neumann coefficient in section 3 and consider the trivial
relativistic solutions for sinh-Gordon and O(N) in section 4.
In the following section we will review the $AdS_5 \times S^5$ elliptic curve and proceed to analyze and solve
the relevant functional equations on the  $AdS_5 \times S^5$ torus. Finally we will describe the pp-wave,
weak coupling and large $L$ limits of the obtained solutions. We close the paper with a discussion and outlook.

\section{String Field Theory vertex axioms}
\label{s.axioms}

\begin{figure}[t]
\hfill\includegraphics[height=3cm]{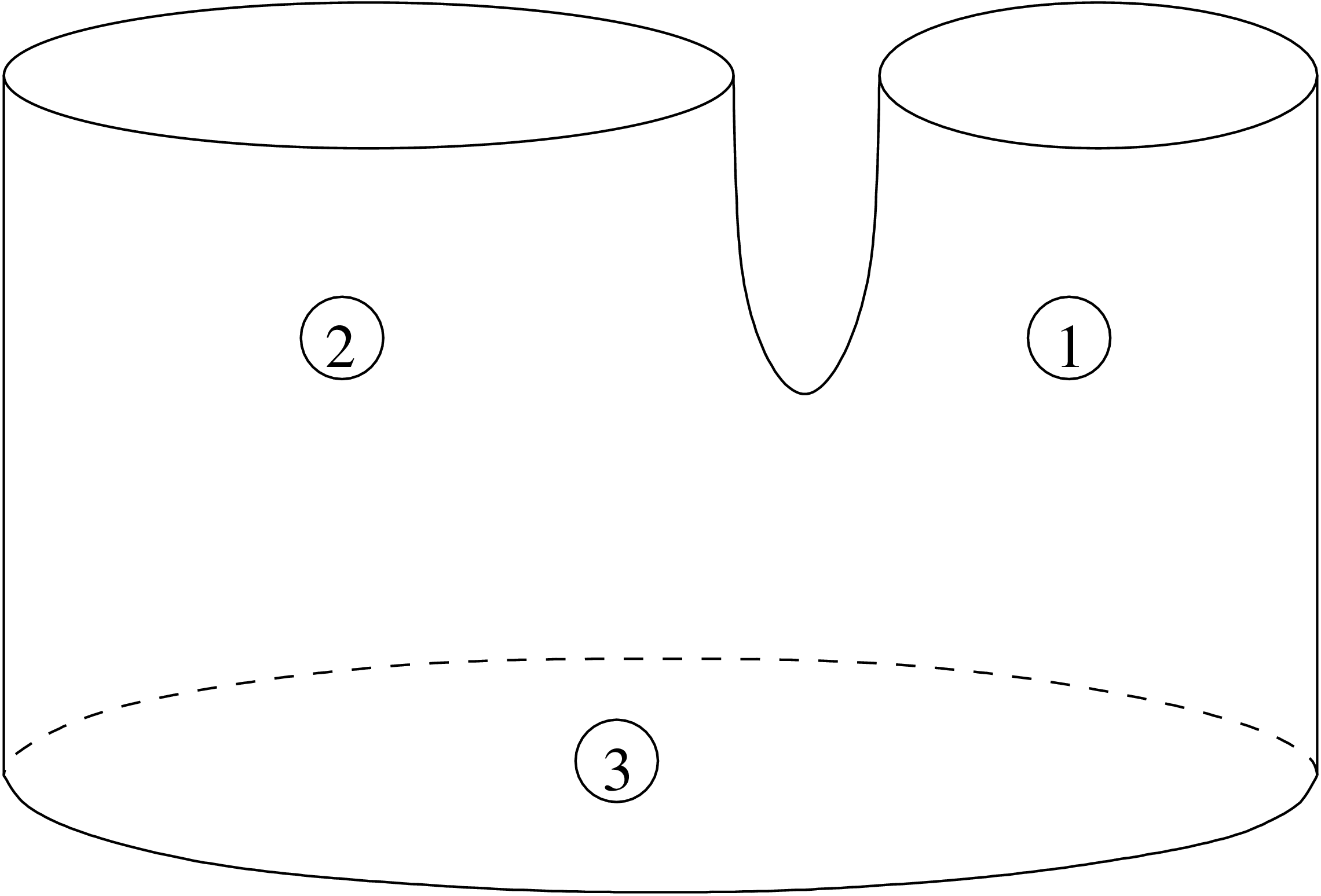}\hfill\includegraphics[height=3cm]{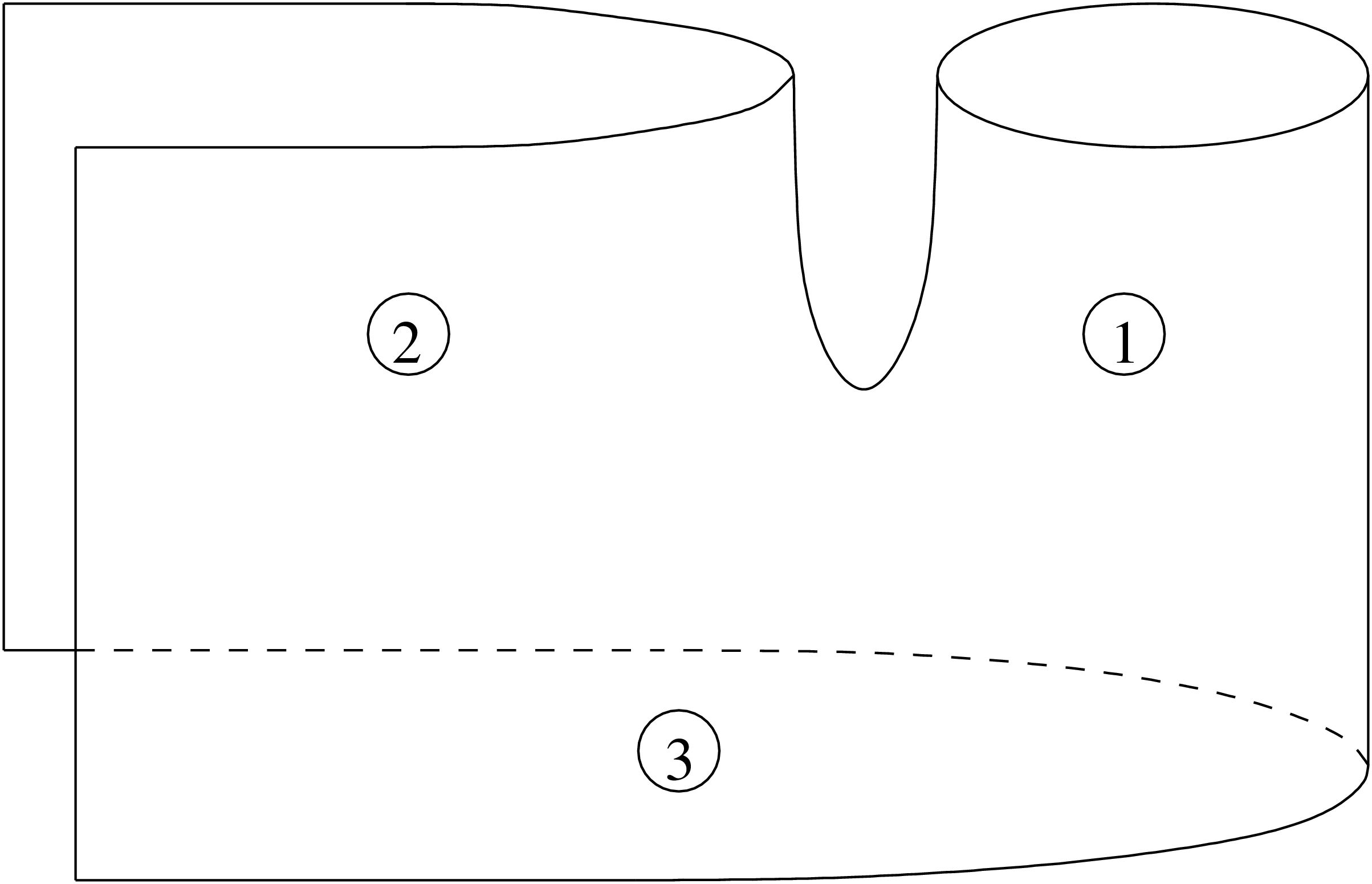}\hfill{}
\caption{The SFT vertex and its decmpactified version.
\label{f.sftpants}}
\end{figure}

The universal exponential part of the light cone string field theory vertex both in flat spacetime and
in the pp-wave geometry has the form
\eq
\label{e.vertex}
\ket{V}=\exp\left\{\f{1}{2} \sum_{r,s=1}^3 \sum_{n,m} N^{rs}_{nm}\, a^{+(r)}_n a^{+(s)}_m \right\} \ket{0}
\eqx
Here $r$ and $s$ labels the three strings in Fig.~\ref{f.sftpants}, $a^{+(r)}_n$ are the corresponding creation operators for
excitations of mode number $n$ on string \#$r$, while the numerical coefficients $N^{rs}_{nm}$ are the
so-called Neumann coefficients. Physically they represent matrix elements of the three string vertex
with just two particles distributed among the three strings.

In the case of \emph{interacting} worldsheet theory, we no longer expect the exponential form (\ref{e.vertex})
to hold, and \emph{a-priori} we will expect to have independent amplitudes for any number of particles\footnote{Of course,
there are some relations between the amplitudes with various numbers of particles, but we do not expect them
to be as simple as following from an exponential form of the vertex.}:
\eq
\Nfin^{3|2;1}_{L_3|L_2;L_1}\left( \th_1,\ldots,\th_n \;\biggl|\; \th'_1,\ldots,\th'_m\, ;\, \th''_1,\ldots,\th''_l \right)
\eqx
As argued in \cite{SFTUS}, we will consider the decompactified vertex with the strings \#2 and \#3 being infinite, and the string \#1 
being of size $L$ (see fig.~\ref{f.sftpants}). 
\eq
\Nfin^{3|2;1}_{\infty|\infty;L}\left( \th_1,\ldots,\th_n \;\biggl|\; \th'_1,\ldots,\th'_m\, ;\, \th''_1,\ldots,\th''_l \right)
\eqx
In this case the functional equations will only depend explicitly on the particles in strings \#2 and \#3, so we can
use a shorthand notation
\eq
N_{\bullet,L}^{3\vert2} \left( \th_1,\ldots,\th_n \;\biggl|\; \th'_1,\ldots,\th'_m \right)
\eqx
where the $\bullet$ stands for a specific state on string \#1: $\bullet \equiv \{ \th''_1,\ldots,\th''_l \}$.

In this paper we will restrict ourselves to amplitudes with just two particles. In analogy to the Minkowski and pp-wave case
we will use the term Neumann coefficients for them. Without loss of generality we can take the two particles to
be in the incoming string \#3. In the notation of \cite{SFTUS},
we have
\eq
N_{\bullet,L}(\theta_{1},\theta_{2})_{i_{1},i_{2}}=N_{\bullet,L}^{3\vert2}(\theta_{1},
\theta_{2}\vert\varnothing)_{i_{1},i_{2}}
\eqx
Also on string \#1 we will put the vacuum state\footnote{The equations for a generic state on string \#1 are identical
but we expect a much more complicated analytical structure with nontrivial additional CDD factors. We leave the investigation
of these interesting and important issues for future work.} $\bullet= \varnothing$.

\subsubsection*{Two particle SFT vertex axioms}

The axioms from \cite{SFTUS} in the case of two particles reduce to
\eqn
\label{e.sft1}
N_{\bullet,L}(\theta_{1},\theta_{2})_{i_{1},i_{2}} &=& S_{i_{1}i_{2}}^{kl}(\theta_{1},\theta_{2})
N_{\bullet,L}(\theta_{2},\theta_{1})_{l,k} \\
N_{\bullet,L}(\theta_{1},\theta_{2})_{i_{1},i_{2}} &=& e^{-ip(\theta_{1})L}N_{\bullet,L}
(\theta_{2},\theta_{1}-2i\pi)_{i_{2},i_{1}} \\
\label{e.sft3}
-i\mbox{Res}_{\theta'=\theta}N_{\bullet,L}(\theta'+i\pi,\theta)_{\bar{i},i} &=&
\left( 1-e^{ip(\theta)L} \right) N_{\bullet,L}
\eqnx
From now on, we will normalize our formulas by setting $N_{\bullet,L}=1$.

Solving these axioms with nontrivial nondiagonal S-matrix does not seem \emph{a-priori} simple,
however in the special case of two particles we can look for a solution of the form
\eq
N_{\bullet,L}(\theta_{1},\theta_{2})_{i_{1},i_{2}} \equiv N(\theta_{1},\theta_{2}) \cdot F(\theta_{1},\theta_{2})_{i_{1},i_{2}}
\label{factoransatz}
\eqx 
where $F(\theta_{1},\theta_{2})_{i_{1},i_{2}}$ satisfies the standard $L$-independent form-factor axioms
\eqn
F(\theta_{1},\theta_{2})_{i_{1},i_{2}} &=& S_{i_{1}i_{2}}^{kl}(\theta_{1},\theta_{2}) F(\theta_{2},\theta_{1})_{l,k} \\
F(\theta_{1},\theta_{2})_{i_{1},i_{2}} &=&  F(\theta_{2},\theta_{1}-2i\pi)_{i_{2},i_{1}}
\eqnx
supplemented with the condition
\eq
\label{e.ffcond}
F(\theta+i\pi,\theta)_{k,i} = \dl_{k\bar{i}} 
\eqx
Then it is easy to show that the two particle SFT axioms (\ref{e.sft1})-(\ref{e.sft3}) will be satisfied
provided that the scalar $N(\theta_{1},\theta_{2})$ satisfies the SFT vertex axioms for a noninteracting theory i.e. with $S=1$:
\eqn
\label{e.nn1}
N(\th_1,\th_2) &=& N(\th_2,\th_1) \\
N(\th_1,\th_2) &=& e^{-ip(\theta_{1})L} N(\th_2,\th_1-2\pi i) \\
-i\mbox{Res}_{\theta'=\theta}N(\theta'+i\pi,\theta) &=& \left( 1-e^{ip(\theta)L} \right)
\label{e.nn3}
\eqnx
For a relativistic theory, these are exactly the axioms satisfied by the (decompactified) pp-wave Neumann coefficients
which are explicitly known. Hence in the relativistic case the problem of finding a solution of the vertex axioms with two particles only reduces
to finding ordinary form factors satisfying the additional condition (\ref{e.ffcond}).

The remaining freedom is a multiplication by a SFT analog of a CDD factor $f(\th_1,\th_2)$ which satisfies
the simple equations
\eq
\label{e.cdd}
f(\th_1,\th_2)=f(\th_2,\th_1) \quad\quad f(\th_1,\th_2)=f(\th_2,\th_1-2\pi i) \quad\quad f(\theta'+i\pi,\theta)=1
\eqx

The goal of this paper is to solve the counterpart of (\ref{e.nn1})-(\ref{e.nn3})
in the case of $AdS_5 \times S^5$ kinematics, where the rapidities live on the appropriate covering space of the torus \cite{CROSS},
and the counterpart of the shift by $i\pi$ is a shift by a half-period of the corresponding elliptic curve.
For obvious reasons we will call the resulting functions \emph{kinematical} Neumann coefficients.

\section{The pp-wave Neumann coefficient}

Before addressing the case of the $AdS_5 \times S^5$ kinematics, let us describe in some detail the (decompactified) pp-wave
Neumann coefficients. Their general structure will also form a guiding principle for seeking a generalization
to the full $AdS_5 \times S^5$ kinematics, as of course the pp-wave relativistic limit can be understood
as a very specific corner in the full $AdS_5 \times S^5$ moduli space at strong coupling.

We are interested here in the $N^{33}(\th_1,\th_2)$ Neumann coefficient which we will denote from now on as 
$N_{pp-wave}(\th_1,\th_2)$. It's decompactified limit can be written in the following
form\footnote{Here we extracted a numerical factor for later convenience.}~\cite{SFTUS}
\eq
\label{e.n33pp}
N_{pp-wave}(\th_1,\th_2) \equiv
N^{33}(\th_1,\th_2)= \f{2\pi^2}{L} \cdot \underbrace{\f{1+\tanh \f{\th_1}{2} \tanh\f{\th_2}{2}}{M\cosh \th_1+ M\cosh\th_2}}_{P(\th_1,\th_2)} n(\th_1) n(\th_2)
\eqx
which will be convenient for generalization to the $AdS_5 \times S^5$ case. Let us first analyze the $P(\th_1,\th_2)$ factor.
It implements for us the kinematical singularity (\ref{e.nn3}). The denominator has a very simple interpretation
as a sum of the energies of the two particles. This will have an obvious generalization to the full $AdS_5 \times S^5$
context, however the drawback of such an expression is that there is an additional spurious singularity at $\th_1=-\th_2+i\pi$ in addition
to the correct kinematical singularity at $\th_1=\th_2+i\pi$. The role of the $\tanh$ functions in the numerator
is exactly to cancel this spurious singularity in a minimal way:
\eq
P(\th_1,\th_2) = \f{1+\tanh \f{\th_1}{2} \tanh\f{\th_2}{2}}{M\cosh \th_1+ M\cosh\th_2} =
\f{\f{1}{\cosh \f{\th_1}{2}} \f{1}{\cosh \f{\th_2}{2}} \cdot  \cosh \f{\th_1+\th_2}{2} }{ 2 M\cosh \f{\th_1-\th_2}{2} 
\cosh \f{\th_1+\th_2}{2} }
\eqx
Since the residue of $P$ at the kinematical pole is 
\eq
\label{e.resp}
-i \mbox{Res}_{\theta'=\theta}P(\theta'+i\pi,\theta) = \f{2 i}{M\sinh \th}
\eqx
and $P(\th_1,\th_2)$ is symmetric and $2\pi i$-periodic, the remaining axioms (\ref{e.nn1})-(\ref{e.nn3}) become
\eqn
\label{e.nmon}
n(\th+2\pi i) &=& e^{-ip(\th)L}\, n(\th) \\
n(\th)\, n(\th+i\pi) &=& \f{1}{2i} \f{L}{2\pi^2} M\sinh\th \left( 1-e^{ip(\theta)L} \right)
\label{e.nkin}
\eqnx
The monodromy relation (\ref{e.nmon}) in fact follows from (\ref{e.nkin}), but it is convenient to first extract a simple
solution of (\ref{e.nmon}) and then deal with a $2\pi i$-periodic function satisfying a modified version of (\ref{e.nkin}).
Namely we introduce
\eq
\label{e.n}
n(\th) = e^{-\f{\th}{2\pi} p(\th) L} \ntl(\th)
\eqx
Then $\ntl(\th)$ is $2\pi i$-periodic and satisfies
\eq
\label{e.ntlkin}
\ntl(\th)\, \ntl(\th+i\pi) = - \f{L}{2\pi^2} M\sinh \th \, \sin \f{p(\th)L}{2}
\eqx
There are many solutions to this equation, but once we require that the zeros lie on the line $\Re e (\th)=0$,
the solution is given by
\eq
\label{e.ntl}
\ntl(\th)=\f{1}{\Gt(\th+i\pi)} \equiv -\f{L}{2\pi^2} M\sinh \th \, \sin \f{p(\th)L}{2} \cdot \Gt(\th)
\eqx
where $\Gt(\th)$ is a new special functions introduced in \cite{LSNS} and slightly redefined in \cite{SFTUS}.
Let us write directly a product representation for $\ntl(\th)$ denoting $\mu=ML/(2\pi)$
\eq
\label{e.ntlpp}
\ntl(\th)=e^{-\mu \cosh \th (\gm+\log \f{\mu}{2e})} \cdot \mu \sinh\th \cdot \prod_{n=1}^\infty \f{\sqrt{n^2+\mu^2} - \mu \cosh\th}{n}
e^{\f{\mu \cosh \th}{n}}
\eqx
The product factors in the numerator ensure that all the nontrivial zeroes required by the rhs of (\ref{e.ntlkin})
lie on the real line and that there are no zeroes on the line $\Re e ( \th) =\pi$. The prefactor, which does not have 
any pole or zero can be understood 
from the large $L$ asymptotics.  Since for large $\mu$ 
\eq
\label{e.ntlas}
\mu \sinh\th \cdot \prod_{n=1}^\infty \f{\sqrt{n^2+\mu^2} - \mu \cosh\th}{n}
e^{\f{\mu \cosh \th}{n}} = - 2 e^{\mu \cosh{\theta} (\gamma +\log \frac{\mu}{2e})} e^{\frac{\theta}{2\pi}pL} 
\sqrt{\frac{\mu}{\pi}} \sin \frac{pL}{2}\cosh \frac{\theta }{2} +O(e^{-\mu}) \,
\eqx
the prefactor simply kills the exponentially large growth of $n(\theta)$. Observe also that in this limit 
the monodromy of $n(\theta)$ is cancelled due to the appearance of the $e^{\frac{\theta}{2\pi}pL}$ factor.
Note, however, that this asymptotics is only valid in the open interval $\Im m(\th) \in (0,2\pi)$ so for
any finite $L$, $\ntl(\th)$ remains a periodic function. We will return to this point later in section~\ref{s.largeL}.

In the next section we will write the solutions for sinh-Gordon and O(N) model and continue 
in the following section to introduce the covering space of the $AdS_5 \times S^5$ torus, describe some general features
of function theory on the elliptic curve and then we will proceed to generalize the structures and formulas encountered
in the present section to the fully general $AdS_5 \times S^5$ case.

\section{Interacting relativistic integrable QFT's}

Before we quote the relevant formulas let us first comment on the meaning of the solutions of the SFT vertex axioms
in the case of such relativistic integrable field theories like sinh-Gordon or O(N) model which clearly 
do not form a consistent string theory. Indeed it is important to note that the SFT vertex axioms from \cite{SFTUS} do not require that. 
They just describe the behaviour
of an integrable quantum field theory on a two-dimensional spacetime which has the geometry of the decompactified pants diagram as in fig.~\ref{f.sftpants}(right).
Clearly we may put \emph{any} quantum field theory on such a geometry and investigate its properties.
This is similar to the question of the spectrum of a QFT on a cylinder which can be formulated for any QFT
without any requirement of a string theory interpretation. 

Let us note in passing that the question of determining what are the properties of an integrable QFT
which ensure that it can arise as a consistent string theory in some gauge-fixing is currently completely 
unexplored.

From the discussion in section \ref{s.axioms} it is clear that the minimal two-particle solutions of the SFT vertex axioms
of any relativistic integrable QFT will have its volume dependence given by the pp-wave Neumann coefficient  $N_{pp-wave}(\th_1,\th_2)$ given by
equations (\ref{e.n33pp}), (\ref{e.n}) and (\ref{e.ntl}). The remaining ingredient is an appropriately normalized minimal form factor
solution.

Thus for sinh-Gordon we have
\eq
N_{min,L}^{shG}(\theta_{1},\theta_{2})= N_{pp-wave}(\th_1,\th_2) \cdot \f{f_{min}^{shG}(\th_1-\th_2)}{f_{min}^{shG}(i\pi)}
\eqx
where $f_{min}^{shG}(\th)$ is the standard sinh-Gordon minimal form factor \cite{SHGFF}
\begin{equation}
f_{min}^{shG}(\th)=\exp\left\{ 4\int\frac{dt}{t}\frac{\sinh\left(t p\right)\sinh
\left(t(1-p)\right)}{\cosh(t)\,\sinh(2t)}\sin^{2}\left(\frac{t}{\pi}(i\pi-\theta)
\right)\right\} 
\end{equation}
where $p$ is related to the sinh-Gordon coupling constant. 

For the O(N) model we have to be slightly more careful and choose the minimal form factor in the singlet channel.
Thus we get
\eq
N_{min,L}^{O(N)}(\theta_{1},\theta_{2})_{i_1 i_2}= N_{pp-wave}(\th_1,\th_2) \cdot 
\f{f_{min}^{singlet}(\th_1-\th_2)}{f_{min}^{singlet}(i\pi)} \dl_{i_1 i_2}
\eqx
where \cite{ONSINGLETFF}
\eq
f_{min}^{singlet}(\th) = \frac{\sinh \theta }{i\pi -\theta } \exp\left\{ 2\int\frac{dt}{t\sinh(t)}\frac{ 1- e^{-t \nu }}{1+ e^{-t}}
\sin^{2}\left(\frac{t}{2\pi}(i\pi-\theta) \right)\right\} 
\eqx
and $\nu = \frac{2}{N-2}$. 

We give these formulas here explicitly as it would be very interesting to cross-check them with some direct
construction of these relativistic integrable QFT's e.g. through some integrable lattice discretization.
This would be important as it would shed light on whether such a minimal solution is indeed the physical one
or whether one should also include some more complicated CDD factors possibly with some additional $L$ dependence.

\section{The $AdS_5 \times S^5$ elliptic curve}

In \cite{CROSS} it was argued that a natural parametrization of the kinematics of a single excitation
of the $AdS_5 \times S^5$ string is given by the universal covering of an appropriate, coupling constant
dependent elliptic curve (equivalently a torus).

Here we will review the relevant formulas as given in \cite{AFREVIEW}, modyfing their definition of $g$ by
a factor of $2$ in order to agree with
\eq
g^2= \f{\lm}{16 \pi^2}
\eqx
so that the dispersion relation is given by
\eq
E=\sqrt{1+16 g^2 \sin^2 \f{p}{2}}
\eqx
The key quantities are $x^\pm$ satisfying
\eq
x^+ + \f{1}{x^+} -x^- - \f{1}{x^-} = \f{i}{g} \qqqq \f{x^+}{x^-} =e^{ip}
\eqx
The modulus of the elliptic curve is $k=-16g^2$, and we have
\eq
2\om_1 = 4K(k) \qqqq 2\om_2=4i K(1-k) -4K(k)
\eqx
where $\om_1$ is related to the periodicity of momentum, while $\om_2$ is the crossing half-period.
Let us also denote by $w$ the relevant complex variable on the universal covering space of the torus. 
Then we have\footnote{$k$ is given in the conventions of {\sl Mathematica}.
From now on we will often suppress giving $k$ explicitly.}
\eq
E=\dn(w,k) \qq \sin \f{p}{2} = \sn(w,k) \qq p=2\am w 
\eqx 
and
\eq
x^\pm = \f{1}{4g} \left( \f{\cn w}{\sn w} \pm i \right) \cdot \left(1+ \dn w \right)
\eqx
Note that the worldsheet momentum $p$ is not globally well defined on the complex plane. 
This will lead to significant complications in solving the SFT vertex axioms
which we will discuss in the next section.

The definitions given above are very concise, however they partly obscure the natural periodicity
as $p \to p+2\pi$ when $w \to w+\om_1$.
Hence we expect that the physics should be described by a torus with periods $\om_1$ and $2\om_2$.

To make this explicit, and also to use $\th$ functions we will often work with the rescaled complex variable
\eq
z =\f{w}{\om_1}
\eqx
and the elliptic curve will have the modular parameter
\eq
\tau = \f{2\om_2}{\om_1}
\eqx
For compatibility with the mathematical definitions that we will be using later, we define 
\eq
q=e^{i \pi \tau}
\eqx
Let us now review the weak coupling and pp-wave limits of the above parametrization.

\subsubsection*{The weak coupling limit}

In the weak coupling limit, the period $\om_1 \to \pi$, while $\om_2 \to i \infty$.
The $z$ coordinate becomes simply related to the worldsheet momentum
\eq
p(z) \sim 2\pi z
\eqx
while the energy becomes
\eq
E(z) \sim 1+8 g^2 \sin^2 \pi z
\eqx

\subsubsection*{The pp-wave limit}

At strong coupling the periods $\om_1$, $\om_2$ have the following expansion
\eq
\om_1 \sim \f{\log g+ 4\log 2}{2g} \qqqq  \om_2 \sim \f{i\pi}{4g}
\eqx
The second formula strongly suggests identifying the relativistic rapidity $\th$
in the pp-wave limit with
\eq
\label{e.ppwaveth}
\th= 4 g w = 4 g \om_1 z
\eqx
Then the crossing transformation is $\th \to \th+i \pi$. One subtlety that one has to keep in mind
is that the pp-wave definition of the momentum $\tilde{p}$ differs from the standard one by an
appropriate rescaling
\eq
\tilde{p}\equiv 2 g p
\eqx
Then indeed $E \to \sqrt{1+\tilde{p}^2}$ and $\tilde{p} = \sinh \th$.
Let us note that due to the behaviour of $\om_1$, after the rescaling (\ref{e.ppwaveth})
the edge of the torus related to momentum periodicity gets pushed to infinity.

\section{Functional equations on the $AdS$ torus}

The functional  equations for the kinematical Neumann coefficients for $AdS_5 \times S^5$ are given by
\eqn
\label{e.nn1el}
N(z_1,z_2) &=& N(z_2,z_1) \\
N(z_1,z_2) &=& e^{-ip(z_{1})L} N(z_2,z_1-\tau) \\
-i\mbox{Res}_{z'=z}\; N\!\left(z'+\f{\tau}{2},z\right) &=& \left( 1-e^{ip(z)L} \right)
\label{e.nn3el}
\eqnx
We will supplement these equations with the requirement that the zeroes of $N(z_1,z_2)$ lie
on the physical line ($\Im m( z)=0$).

Despite their structural similarity with the relativistic equations, the highly rigid
function theory on a torus leads to various stringent restrictions and puzzles.
In particular the worldsheet momentum $p$ is not globally well defined on the complex plane.
This has two consequences. Firstly, the exponential factors $e^{ipL}$ are much more heavily constrained than in the relativistic
case. Indeed they are well defined meromorphic functions only for integer $L$ (for half integer $L$
they are also meromorphic but on a larger torus with periodicity $z \to z+2$). This property is indeed very natural from the gauge
theory point of view as the size of the string is always integer (or half-integer) as it is identified with the discrete $J$ charge.

This new feature of the AdS kinematics will also severly complicate solving the SFT vertex monodromy axiom.
Indeed  a function of the form
\eq
e^{const \cdot z \cdot p(z)}
\eqx
similar to the function $e^{-\f{\th}{2\pi} p(\th) L}$ which was used in the relativistic case in (\ref{e.n}) does not make sense
on the elliptic curve (or on its covering space) as it has branch cuts and is not meromorphic.

Let us now turn to finding a solution of (\ref{e.nn1el})-(\ref{e.nn3el}). 
Instead of directly attacking the functional relations (\ref{e.nn1el})-(\ref{e.nn3el}), we will try to follow
the steps employed when solving the functional relations in the pp-wave case, and decompose $N(z_1,z_2)$ into
some simpler structures.
Recall~(\ref{e.n33pp}):
\eq
N^{33}(\th_1,\th_2)= \f{2\pi^2}{L} \cdot \underbrace{\f{1+\tanh \f{\th_1}{2} \tanh\f{\th_2}{2}}{M\cosh \th_1+ M\cosh\th_2}}_{P(\th_1,\th_2)} n(\th_1) n(\th_2)
\eqx
We will look for a similar decomposition
\eq
N(z_1,z_2) = \f{2\pi^2}{L} \cdot \underbrace{\f{1+f(z_1) f(z_2)}{E(z_1)+ E(z_2)}}_{P(z_1,z_2)} n(z_1) n(z_2)
\eqx
with the functions $f(z)$ and $n(z)$ to be determined.

\subsection*{The function $f(z)$}

The key role of the numerator in $P(z_1,z_2)$ is to cancel the unwanted pole at $z_1=-z_2+\tau/2$ in
the denominator. Since we want the solution to reduce to the pp-wave solution in the appropriate limit, we will make a shortcut
and try to find a natural elliptic generalization of $\tanh \th/2$. The key properties of $\tanh \th/2$
which are also necessary to cancel that spurious pole amount to
\eq
f(z+\tau/2) = \f{1}{f(z)} \qqqq f(-z)=-f(z)
\eqx

Elliptic functions can be constructed in diverse ways. For later convenience we will use the $q$-theta function $\th_0(z)$
defined through
\eq
\th_0(z)=-i e^{i\pi \left(z-\f{\tau}{4}\right) +i\pi \f{\tau}{12}} \cdot \f{\th_1(\pi z, e^{i\pi \tau})}{\eta(\tau)}
\eqx
as a basic building block. This function obeys the properties:
\eqn
\th_0(z+1) &=& \th_0(z) \\
\th_0(z+\tau) &=& -e^{-2\pi i z} \th_0(z) \\
\th_0(\tau-z) &=& \th_0(z) \\
\th_0(-z) &=& -e^{-2\pi i z} \th_0(z)
\eqnx
Its main property is that it has a single zero at $z=0$. Thus it may be used to construct elliptic functions
by specifying the positions of their zeroes and poles. Indeed any elliptic function can be written
as
\eq
const \cdot \f{ \th_0(z-a_1) \th_0(z-a_2) \cdot \ldots \cdot \th_0(z-a_n) }{ \th_0(z-b_1) \th_0(z-b_2) \cdot \ldots \cdot \th_0(z-b_n) }
\eqx
with the constraint $\sum_{i=1}^n  a_i = \sum_{i=1}^n b_i$ for double periodicity. It is well known that
the elliptic functions have to have $n \geq 2$.

\begin{figure}[t]
\hfill\includegraphics[height=6cm]{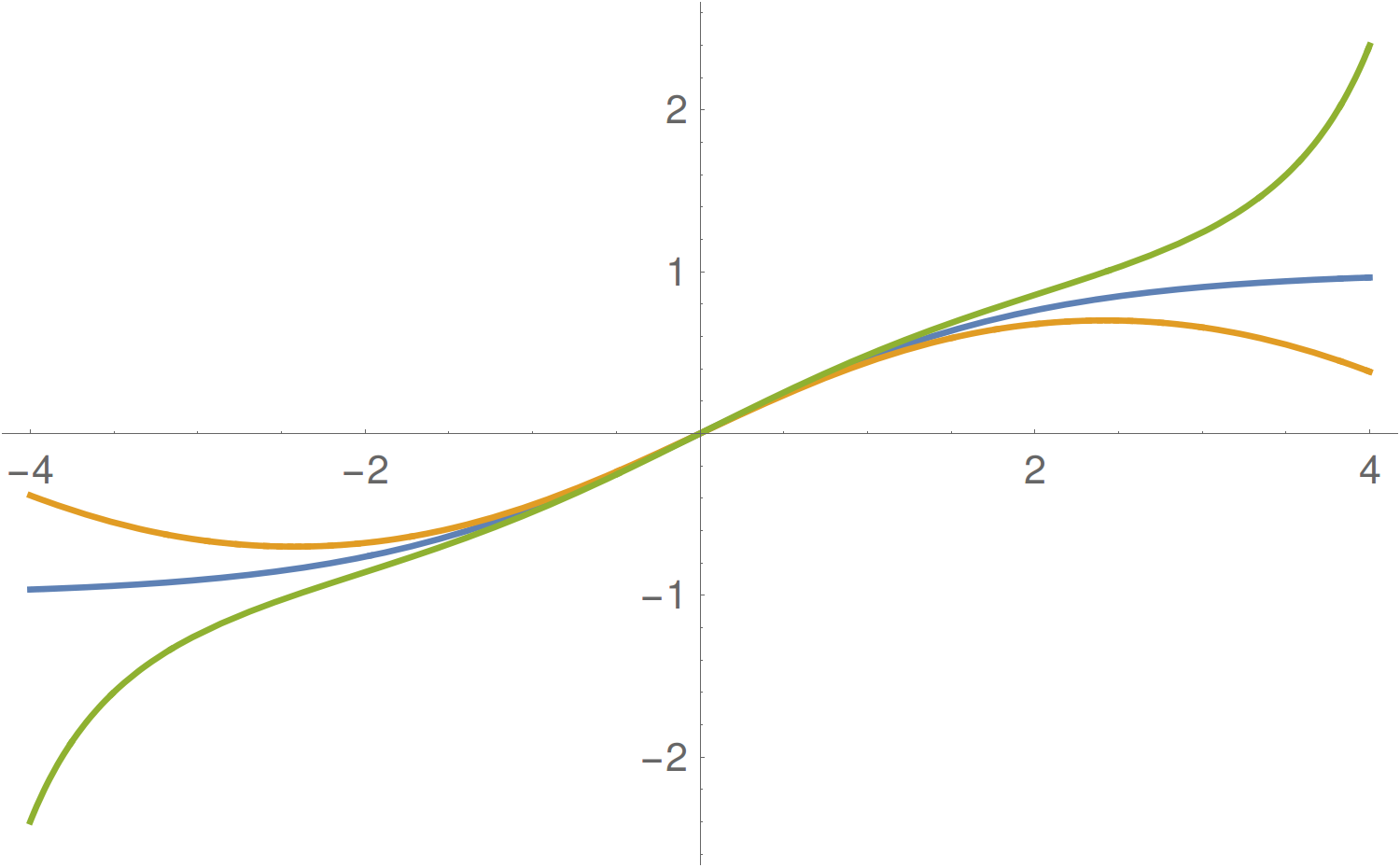}\hfill{}
\caption{The functions $f(z)$ with the two choices of $z_0$ and $\tanh \f{\th}{2}$ for $\lm=10000$.   
\label{f.tanh}}
\end{figure}

The function $f(z)$ thus has to have the following form
\eq
\label{e.fz0}
f(z)=C \f{ \qth{z} \qth{z-z_0} }{ \qth{z-\f{\tau}{2}} \qth{z-z_0+ \f{\tau}{2}} }
\eqx
In order for $f(z)f(z+\tau/2)=1$ to hold, $C$ can be calculated to be
\eq
C=\mp e^{i \pi \left( z_0 -\f{\tau}{2} \right)}
\eqx
In the following we will pick the upper sign.
In order for this function to be odd, $z_0$ has to be a half period. We have two possibilities:
\eq
z_0=\f{1}{2}  \qq \text{or} \qq  z_0=\f{1+\tau}{2}
\eqx
Provisionally we will use the function with the first choice of $z_0$ 
as it has no pole on the physical line. Thus we set
\eq
f(z) = - i q^{-\f{1}{2}} \cdot \f{ \qth{z} \qth{z-\f{1}{2}} }{ \qth{z-\f{\tau}{2}} \qth{z-\f{1}{2}+ \f{\tau}{2}} }
\eqx
Both choices of $z_0$ in (\ref{e.fz0}), however, lead to functions which go over to $\tanh \f{\th}{2}$ in the pp-wave limit as can be seen in figure~\ref{f.tanh}

\subsection*{The function $n(z)$}

Let us now consider the analog of (\ref{e.resp}) which for the reader's convenience we repeat here
\eq
-i \mbox{Res}_{\theta'=\theta}P(\theta'+i\pi,\theta) = \f{2 i}{M\sinh \th}
\eqx
It would be tempting to identify the expression $M\sinh\th$ in the residue with the momentum, but on
the elliptic curve this would be problematic, as the momentum is not a well defined function.
In fact it can be equivalently understood as $E'(\th)$, which in contrast has a well defined
elliptic generalization. Since in the previous subsection we have already explicitly defined
$P(z_1,z_2)$, of course we do not have any freedom here but we just have to compute the appropriate residue.
It is quite encouraging that $E'(z)$ indeed appears in the exact answer:
\eq
-i \mbox{Res}_{z'=z}P(z'+\tau/2,z) = \f{2 i}{E'(z)}
\eqx
We are now left with the following functional equations for $n(z)$:
\eqn
n(z+\tau) &=& e^{-i p(z)L} n(z) \\
n(z)n(z+\tau/2) &=& \f{L E'(z)}{4\pi^2 i}  \left( 1-e^{ip(z)L} \right)
\label{e.second}
\eqnx
Again the first equation is a direct consequence of the second one.
For later convenience let us give an expression for $E'(z)$ in terms
of the momenta:
\eq
E'(z) =-4 g^2 \om_1 i \left(e^{ip}- e^{-ip}\right)= 8 g^2 \sin p
\eqx
The relevant crossing equation for $n(z)$ becomes then
\eq
n(z)n(z+\tau/2) = -\f{4 g^2 L}{\pi^2} \sin p \sin \f{pL}{2} e^{\f{ipL}{2}}
\eqx

In this paper we will concentrate on the case of even $L=2n$ which is simpler than the general case.
Let us first construct an \emph{elliptic function} $G_L^{ev}(z)$ which has the correct location
of zeroes following from (\ref{e.second}). Then we will concentrate on solving the monodromy equation in the 
simplest possible setting.
Similarly as in the pp-wave limit,
we will require all the zeroes (in the fundamental domain) to lie on the physical real axis.

It is natural to implement this condition by defining
\eq
G_{L=2n}^{ev}(z) = \sqrt{\f{L}{2}} \prod_{k=1}^{n-1} \f{  \sqrt{1+16g^2 \sin^2 \f{\pi k}{L}}- E(z) }{4 g \sin \f{\pi k}{L}}
\eqx
This function satisfies the following functional equation
\eq
G_L^{ev}(z) G_L^{ev}(z+\tau/2)= \f{\sin \f{pL}{2}}{\sin p}
\eqx
Let us now write $n(z)$ as
\eq
n(z)=\f{2g \sqrt{L}}{\pi} \sin p\; G_L^{ev}(z) h^{ev}_L(z)
\eqx
Then the remaining function $h^{ev}_L(z)$ will satisfy a very simple equation
\eq
\label{e.hevproduct}
h^{ev}_L(z) h^{ev}_L(z+\tau/2) = e^{\f{ipL}{2}}
\eqx
leading to
\eq
h^{ev}_L(z+\tau)= e^{-ipL} h^{ev}_L(z)
\eqx
This function will be the direct elliptic counterpart of $e^{-\f{\th}{2\pi} p(\th) L}$ in the relativistic case,
however the analyticity properties in the `elliptic' rapidity plane force the solution to be much
more complicated.

\subsection{Elliptic Gamma function and the monodromy condition}

In order to solve the monodromy functional equations we will need to use the so-called elliptic Gamma function $\Gm(z,\tau,\sg)$.
Its definition and main properties are discussed in \cite{FELDER}.
It is the unique meromorphic solution of the difference equation
\eq
\Gm(z+\sg,\tau,\sg)=\th_0(z,\tau) \Gm(z,\tau,\sg)
\eqx 
such that i) $\Gm(z+1,\tau,\sg)=\Gm(z,\tau,\sg)$, ii) $\Gm(z,\tau,\sg)$ is holomorphic on the upper half plane, and 
it is normalized by iii) $\Gm((\tau+\sg)/2,\tau,\sg)=1$.
It is given by an explicit product representation
\eq
\Gm(z,\tau,\sg) = \prod_{j,k=0}^\infty \f{1-e^{2\pi i ((j+1)\tau+ (k+1)\sg-z)}}{1-e^{2\pi i (j \tau+ k\sg+z)}}
\eqx
In the fundamental domain there are no zeroes and the only poles are on the real line at integer values of $z$.
All other poles occur in the lower half plane. 
In the present paper we will need just the special case with $\sg=\tau$, which we will denote by the shorthand notation
\eq
\Gmel(z) \equiv \Gm(z,\tau,\tau) = \prod_{k=0}^\infty \left( \f{1-e^{2\pi i \tau(k+2)} e^{-2\pi i z}}{1-e^{2\pi i \tau k} e^{2\pi i z}} \right)^{k+1}
\eqx
It satisfies
\eq
\Gmel(z+1)=\Gmel(z) \qq \Gmel(z+\tau)=\th_0(z) \Gmel(z)
\eqx

\subsubsection*{The monodromy condition}

The function $h^{ev}_L(z)$ satisfies the following monodromy condition
\eq
\label{e.hevmonodromy}
h^{ev}_L(z+\tau)= e^{-ipL} h^{ev}_L(z)
\eqx
Let us first investigate the more elementary equation
\eq
\label{e.Hmonodromy}
H(z+\tau)= e^{-ip} H(z)
\eqx
We can readily construct such a function using the elliptic Gamma functions $\Gmel(z)$ once we express
$e^{-ip}$ in terms of the elementary $\th_0$ functions:
\eq
e^{-ip}= q^{\f{1}{2}}  \cdot \f{ \th_0\left(z-\f{1}{2}+\f{\tau}{4}\right) \th_0\left(z-\f{1}{2}-\f{3\tau}{4}\right)}{\th_0^2\left(z-\f{1}{2}-\f{\tau}{4}\right)}
\eqx
Thus the function $H(z)$ satisfying (\ref{e.Hmonodromy}) can be given by
\eq
H(z)= e^{i \f{\pi}{2} z} \cdot
\f{ \Gmel\left(z-\f{1}{2}+\f{\tau}{4}\right) \Gmel\left(z-\f{1}{2}-\f{3\tau}{4}\right) }{ \Gmel^2\left(z-\f{1}{2}-\f{\tau}{4}\right)}
\eqx
However due to the innocous looking leftover constant $q^{\f{1}{2}}$ appearing in the expression for $e^{-ip}$,
we are forced to include the exponential factor $e^{i \f{\pi}{2} z}$ which violates the $z\to z+1$ periodicity.
Indeed $H(z)$ satisfies
\eq
H(z+1)=i H(z)
\eqx
Nevertheless for the case of even $L$ which we are considering in the present paper we may easily obtain 
a $z\to z+1$ periodic solution to (\ref{e.hevmonodromy}).
Let us take first $L=2$. Then the solution is
\eq
C \cdot e^{-i\f{p(z)}{2}} H^2(z) e^{-2ip(z)}
\eqx
The term $e^{-ip/2}$ restores $z\to z+1$ periodicity, the other factor of $e^{-2ip}$ ensures that the expression is real
on the physical line $\Im m(z)=0$, while the constant $C=1/(H(0) H(\tau/2))$ is enough to satisfy the remaining
equation (\ref{e.hevproduct}) for $L=2$. The generalization to any even $L=2n$ is now trivial:
\eq
h^{ev}_{L=2n}(z) = \f{1}{H^n(0) H^n(\tau/2)} \cdot e^{-i \f{p}{2}n} e^{-i p L} H(z)^L
\eqx
This solves all the required equations and is real on the real axis. In the next section we will put all these partial formulas
together and explore some of the properties of the AdS kinematical Neumann coefficient.

\section{The kinematical $AdS_5 \times S^5$ Neumann coefficient}

Let us now collect together the relevant formulas. The resulting expression is an exact solution of the AdS axioms for the kinematical
Neumann coefficient (\ref{e.nn1el})-(\ref{e.nn3el}) for any even value of $L=2n$.
The solution is of course valid for any value of the gauge theory coupling constant.
We have
\eq
N(z_1,z_2) = \f{2\pi^2}{L} \cdot \f{1+f(z_1) f(z_2)}{E(z_1)+ E(z_2)} n(z_1) n(z_2)
\eqx
where
\eq
f(z) = - i q^{-\f{1}{2}} \cdot \f{ \qth{z} \qth{z-\f{1}{2}} }{ \qth{z-\f{\tau}{2}} \qth{z-\f{1}{2}+ \f{\tau}{2}} }
\eqx
while $n(z)$ is composed of two pieces
\eq
n(z)=\f{2g \sqrt{L}}{\pi} \sin p\; G_L^{ev}(z) h^{ev}_L(z)
\eqx
with $G_L^{ev}(z)$ being an elliptic function ensuring the correct positions of zeroes as required by the kinematical singularity axiom
\eq
G_{L=2n}^{ev}(z) = \sqrt{\f{L}{2}} \prod_{k=1}^{n-1} \f{  \sqrt{1+16g^2 \sin^2 \f{\pi k}{L}}- E(z) }{4 g \sin \f{\pi k}{L}}
\eqx
while $h^{ev}_L(z)$ implements the correct monodromy under the shift $z \to z+\tau$
\eq
h^{ev}_{L=2n}(z) = \f{1}{H^n(0) H^n(\tau/2)} \cdot e^{-i \f{p}{2}n} e^{-i p L} H(z)^L
\eqx
with
\eq
H(z)= e^{i \f{\pi}{2} z} \cdot
\f{ \Gmel\left(z-\f{1}{2}+\f{\tau}{4}\right) \Gmel\left(z-\f{1}{2}-\f{3\tau}{4}\right) }{ \Gmel^2\left(z-\f{1}{2}-\f{\tau}{4}\right)}
\eqx
In the following we will discuss the singularitites of the kinematical Neumann coefficient and its pp-wave, weak coupling and large $L$ limits.

\subsection{Singularity structure}

Let us now analyze the singularity structure of the solution $N(z,z')$ as a function of $z$ keeping $z'$ fixed.

From the definition of the elliptic Gamma function we see that the potential zeroes and poles of $h^{ev}_{L=2n}(z)$ can occur only for
the points
\eq
z_1=\f{1}{2}+\f{\tau}{4}  \qqqq z_2=\f{1}{2}+\f{3\tau}{4}
\eqx
in the `fundamental' domain (which we define here as the set $0\leq \Re e(z) <1$, $0\leq \Im m(z) <\tau$).
These points represent the poles and zeroes of $e^{-ip}$ and thus represent infinite (complex) momentum
thus having singularities there is quite natural.

The function $h^{ev}_{L=2n}(z)$ has a pole of order $n$ at $z_1$ and a zero of order $3n$ at $z_2$. The fact that the number of
poles and zeroes does not balance is not a contradiction as this function has nontrivial monodromy in the $\tau$ direction and thus is not elliptic.
The poles of $G_L^{ev}(z)$ just follow from the poles of the energy $E(z)$ which has first order poles both at $z_1$ and $z_2$. Consequently
$G_L^{ev}(z)$ has poles of  order $n-1$ both at $z_1$ and at $z_2$. It also has $2n-2$ zeroes on the real axis (within the `fundamental' domain).
Thus the product of the two functions has a pole of order $2n-1$ at $z_1$ and a zero of order $2n+1$ at $z_2$.
Finally $\sin p$ has poles of order $2$ both at $z_1$ and at $z_2$ and zeroes at $z=0,1/2,\tau/2,1/2+\tau/2$.
Of these zeroes the first two on the real axis are expected, while we will have to track the ones at $z=\tau/2,1/2+\tau/2$.

Therefore $n(z)$ has a pole of order $2n+1$ at $z_1$, a zero of order $2n-1$ at $z_2$ and two single zeroes at  $z=\tau/2,1/2+\tau/2$
apart from the expected set of real zeroes.

It remains to analyze the singularities of the $L$ independent piece
\[
\f{1+f(z) f(z')}{E(z)+ E(z')}
\]
as a function of $z$ (keeping $z'$ fixed). Generically we would expect this function to be an elliptic function of order 4,
but since by construction $f(z)$ was choosen to cancel the unphysical pole at $z=-z'+\tau/2$ it is a function of order 3.
This function has the kinematical pole at $z=z'+\tau/2$, the remaining two first order poles are at  $z=\tau/2,1/2+\tau/2$
which exactly cancel with the complex zeros of $n(z)$. This cancellation is a nice consistency check of this solution.
All the zeroes are of first order. Two of them are at $z_1$ and $z_2$, while the last one is at $z=z'+1/2+\tau/2$.

Putting all these considerations together, we see that the solution $N(z,z')$ has a pole of order $L$ at $z_1$, a zero of order $L$
at $z_2$, a first order pole at  $z=z'+\tau/2$ (the kinematical pole), a set of zeroes on the real axis and an additional
zero at $z=z'+1/2+\tau/2$. It would be interesting to understand the meaning of this additional zero.

Let us just mention in passing that if we define
\eq
N_{reg}(z,z') \equiv N(z,z')\, e^{ip(z)L/2} e^{ip(z')L/2}
\eqx
we can get rid of any zeroes and poles at $z_1$ and $z_2$ altogether.

\subsection{The pp-wave limit}

In order to study the pp-wave limit, we have to take $g\to \infty$ together with $L \to \infty$ keeping fixed
\eq
\tilde{p}=2 g p \qqqq  \tilde{L}=\f{L}{2g}
\eqx
Then, as mentioned earlier, the dispersion relation becomes $E=\sqrt{1+\tilde{p}^2}$ and the relativistic rapidity
is linked with the $z$ coordinate on the torus through
\eq
\th= 4 g \om_1 z
\eqx
Taking this limit analytically on the kinematical Neumann coefficient is rather involved and we did not carry it out in full
but we performed a numerical check. However let us comment first on some partial analytical results which indicate
that the various functions which we introduced like the elliptic function $G_L^{ev}(z)$ and the $h^{ev}_{L=2n}(z)$ containing 
the elliptic Gamma function are in fact strongly interrelated.

\begin{figure}[t]
\hfill\includegraphics[height=6cm]{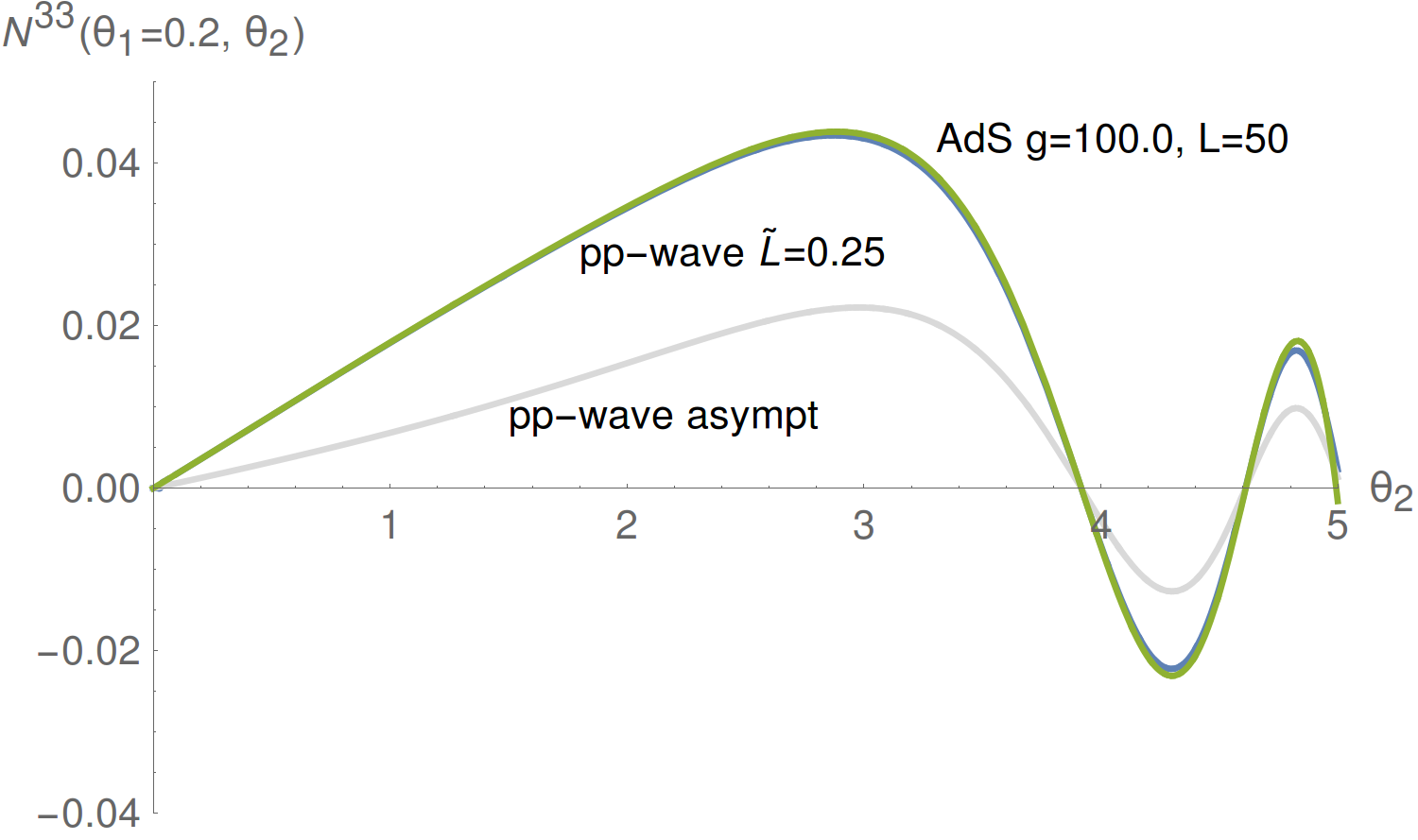}\hfill{}
\caption{The pp-wave Neumann coefficient $N_{pp-wave}(\th_1=0.2,\th_2)$ for $\tilde{L}=0.25$ together with its
asymptotic part (neglecting wrapping) and the full $AdS_5 \times S^5$ kinematical Neumann coefficient
for $g=100$ and $L=50$ (which corresponds to $\tilde{L}=0.25$).
\label{f.ppwave}}
\end{figure}

Using the properties of the elliptic Gamma function in \cite{FELDER} one can obtain the pp-wave limit of $h^{ev}_{L=2n}(z)$:
\eq
h^{ev}_{L}(z) \to e^{-\f{1}{2\pi} \tilde{L}\, \th\, \sinh\th} \cdot e^{\f{1+4\log 2+ \log g}{2\pi}\, \tilde{L}\, \cosh\th}
\eqx
The first term is exactly the relativistic monodromy function used in (\ref{e.n}). The second term, however, involves already
a part of the exponential factor in (\ref{e.ntlpp}), but due to the $\log g$ in the exponent, this function does not
really have a pp-wave limit. It turns out that only when multiplied by $G_L^{ev}(z)$, the $\log g$ term apparently cancels
and we have a well defined pp-wave limit of the complete expression. We checked this numerically (see fig.~\ref{f.ppwave}).
There we compare the full $AdS_5 \times S^5$ answer with the pp-wave expression in the far from asymptotic regime where 
wrapping is important and the full pp-wave exact expression (\ref{e.ntlpp}) is needed.

\section{Weak coupling limit }

In this section we analyze the weak coupling limit $(g\to0)$ of the
kinematical Neumann coefficient and connect it to a decompactified
spin chain calculation. In this limit the real period of the torus,
$\omega_{1}$ goes to $\pi$, while the imaginary one diverges as $\omega_{2}\to i\infty$.
This makes the domains,  related for finite $g$ by crossing, disconnected
at weak coupling. Since the spin chain calculation gives nonvanishing
result only for the kinematics when there is one incoming particle
in string \#3 and one outgoing particle in string \#2 we have to continue
analytically the kinematical Neumann  
coefficient to describe this process  
\eq
\label{e.n23}
N_{23}(z',z) \equiv e^{-ip(z')\f{L}{2}} \, N(z'+\frac{\tau}{2},z)
\eqx 
before taking the weak coupling limit \cite{SFTUS}.
Using the functional relations
\begin{equation}
n(z)=\frac{1}{n(z+\frac{\tau}{2})}\frac{LE'(z)}{4\pi^{2}i}(1-e^{ip(z)L})\quad;
\qquad f(z+\frac{\tau}{2})=\frac{1}{f(z)}
\end{equation}
we transform the required quantity into:
\begin{equation}
N(z'+\frac{\tau}{2},z)=\frac{E'(z)}{2i}\frac{1+\frac{f(z'+\frac{\tau}{2})}{f(z+
\frac{\tau}{2})}}{E(z)-E(z')}\frac{n(z'+\frac{\tau}{2})}{n(z+\frac{\tau}{2})}(1-e^{ipL})
\end{equation}
In  evaluating its weak coupling limit we note that  the
elliptic nome  goes to zero, $q\to0$, and the theta functions simplify to
trigonometric functions. In particular we find that
\begin{equation}
\frac{f(z'+\frac{\tau}{2})}{f(z+\frac{\tau}{2})}=\frac{\sin2\pi z}{\sin2\pi z'}=
\frac{\sin p}{\sin p'}
\end{equation}
where  in the second equality we used that  at weak coupling $p=2\pi z.$  Consequently in this limit we also have
\begin{equation}
E(z)=1+8g^{2}\sin^{2}\pi z+\dots \quad\quad \text{and} \quad\quad E'(z)=8 g^2 \sin p+ \ldots
\end{equation}
which allows us to evaluate the weak coupling limit of
the $L$-independent prefactor.

Let us now turn to analyze
\begin{equation}
\frac{n(z'+\frac{\tau}{2})}{n(z+\frac{\tau}{2})}=\frac{\sin p'}{\sin p}
\frac{G_{L}^{ev}(z'+\frac{\tau}{2})}{G_{L}^{ev}(z+\frac{\tau}{2})}
\frac{h_{L}^{ev}(z'+\frac{\tau}{2})}{h_{L}^{ev}(z+\frac{\tau}{2})}
\end{equation}
Firstly we see that
\begin{equation}
\frac{G_{L}^{ev}(z'+\frac{\tau}{2})}{G_{L}^{ev}(z+\frac{\tau}{2})}=
\prod_{k=1}^{n-1}\frac{\sqrt{1+16g^{2}\sin^{2}\frac{\pi k}{L}}+E(z')}
{\sqrt{1+16g^{2}\sin^{2}\frac{\pi k}{L}}+E(z)}=1+\dots
\end{equation}
The small $q$ limit of the elliptic gamma function comes from the
first factor in the product
\begin{equation}
\Gmel(z)=\prod_{k=0}^{\infty}\left(\frac{1-q^{2(k+2)}e^{-2i\pi z}}
{1-q^{2k}e^{2i\pi z}}\right)^{k+1}=\frac{1}{(1-e^{2i\pi z})}+\dots
\end{equation}
whenever $\Im m(z)>-\tau$. This implies 
\begin{equation}
H(z+\frac{\tau}{2})=q^{\frac{1}{4}}e^{i\frac{\pi}{2}z}\frac{
\Gmel(z-\frac{1}{2}+\frac{3\tau}{4})
\Gmel(z-\frac{1}{2}-\frac{\tau}{4})}{
\Gmel(z-\frac{1}{2}+\frac{\tau}{4})^{2}}=
q^{\frac{3}{4}}e^{-i\frac{3\pi}{2}z}+\dots
\end{equation}
and leads to 
\begin{equation}
\frac{n(z'+\frac{\tau}{2})}{n(z+\frac{\tau}{2})}=\frac{\sin p'}{\sin p}
\frac{e^{2i\pi p'n}}{e^{2i\pi pn}}=\frac{\sin p'}{\sin p}e^{i\frac{L}{2}(p'-p)}
\end{equation}
Putting everything together we obtain 
\begin{equation}
N(z'+\frac{\tau}{2},z)=\frac{\pi}{i}\cot \f{p'-p}{2} e^{i\frac{L}{2}(p'-p)}(1-e^{ipL})
\end{equation}
This implies for the weak coupling limit of the amplitude $N_{23}(z',z)  $:
\eq
N_{23}(z',z) =\pi \f{1+e^{i(p'-p)}}{1-e^{i(p'-p)}} \cdot (1-e^{ipL}) \cdot e^{-i\f{p L}{2}}
\label{weakN23}
\eqx
In the following we compare this result to an infinite volume spin-chain calculation.

\subsection{Decompactifed spin chain calculation}

\begin{figure}
\begin{centering}
\includegraphics[width=8cm]{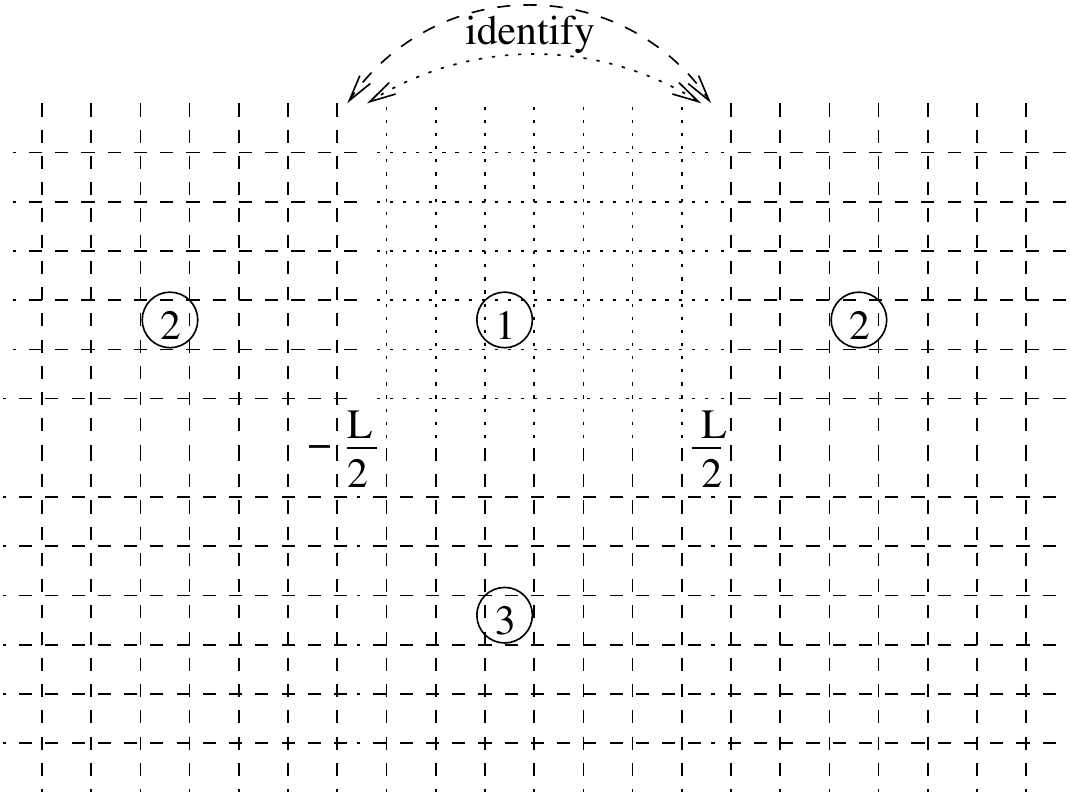}
\par
\end{centering}
\caption{The geometry of the spin chain calculation. The world sheet of string \#3 is replaced 
with an infinite spin chain, which splits into the infinite spin chain of string \#2 and the 
periodic spin chain of size $L$ replacing string \#1.\label{figspinchain}}
\end{figure}

In the weak coupling limit the Neumann coefficients are expected to
be related to the tree level 3pt functions of the dual gauge theory.
To calculate this 3pt functions one has to diagonalize the 1-loop
dilatation operator and evaluate the overlap of its eigenstates in
the decompactified geometry shown on figure \ref{figspinchain}. This
is equivalent to a decompactified spin chain calculation. 
Figure \ref{figspinchain} depicts the geometry in which the decompactified
string \#3 splits into the decompactified string \#2 and the finite
string \#1. We assume that we have one particle for string \#3 with
momentum $p$ and one particle for string \#2 with momentum $p'$,
and the vacuum for string \#1. In the language of gauge theory this
setting translates into three operators, $\mathcal{O}_{i}$, for $i=1,2,3$
as follows: for $\mathcal{O}_{3}$ we take an infinitely long operator
built up from one single $X$ and infinitely many $Z$ scalar operators.
The coordinate space eigenstate of the dilatation operator can be
parameterized by its momentum $p$: 
\begin{equation}
\vert\mathcal{O}_{3}\rangle=\sum_{n\in Z}e^{ipn}\vert n\rangle,
\end{equation}
where $\vert n\rangle$ is of the form $\dots ZZZXZZZ.\dots$ and
the operator $X$ is located at position $n$.%
\footnote{ This state is normalized to $\delta$ function in 
$p$.}. The operator we take for $\mathcal{O}_{2}$ is in the conjugate sector
to $\mathcal{O}_{3}$, it contains infinitely many $\bar{Z}$ and
one single $\bar{X}$. Finally for the third operator we take $\mathcal{O}_{1}=
\mbox{Tr}(\bar{Z}^{L})$,
whose state is $\langle\mathcal{O}_{1}\vert={}_{L}\langle0\vert$.
To implement the right geometry we split $\mathcal{O}_{2}$ as
\begin{equation}
\langle\mathcal{O}_{2}\vert=\sum_{n'\leq-\frac{L}{2}}e^{-ip'(n'+\frac{L}{2})}\langle n'\vert+
\sum_{n'>\frac{L}{2}}e^{-ip'(n'-\frac{L}{2})}\langle n'\vert
\end{equation}
and insert $\mathcal{O}_{1}$ in the middle. This basis is very similar to the one, which was used to calculate 
the decompactified Neumann coefficients in  \cite{SFTUS}. In calculating the overlap
$(\langle\mathcal{O}_{2}\vert\otimes\langle\mathcal{O}_{1}\vert)\vert\mathcal{O}_{3}
\rangle$
we note that at tree level the nontrivial contractions are $\langle n'\vert n\rangle=
\delta_{n,n'}.$
This implies
\begin{eqnarray}
(\langle\mathcal{O}_{2}\vert\otimes\langle\mathcal{O}_{1}\vert)\vert\mathcal{O}_{3}
\rangle & = & e^{-ip'\frac{L}{2}}\sum_{n\leq -\frac{L}{2}}e^{i(p-p')n}+e^{ip'\frac{L}{2}}\sum_{n>\frac{L}{2}}e^{i(p-p')n}\nonumber \\
 & = & \frac{1}{1-e^{i(p'-p)}}  \cdot (1-e^{ipL}) \cdot e^{-i p \frac{L}{2}}
\end{eqnarray}
The above equation is very similar to the one obtained from the weak coupling limit
of the $AdS_5 \times S^5$ kinematical Neumann coefficient (\ref{weakN23}), except the factor $ 1+e^{i(p'-p)} $
which, however,  satisfies the AdS version of the CDD axioms\footnote{The CDD equations
were written for two incoming particles so again we have to cross back. 
In particular this will change $ 1+e^{i(p'-p)} $ to $ 1+e^{-i(p'+p)} $.} (\ref{e.cdd}).
The appearance of such an additional factor is very natural as we factored out the S-matrix 
dependent ordinary two particle form factor (\ref{factoransatz}), which varies from sector to sector. Here, however we 
calculated only the one related to the $su(2)$ sector.

\section{The large $L$ limit}
\label{s.largeL}

It is also interesting to analyze the kinematical Neumann coefficient in the limit of large $L$
keeping the remaining variables like the gauge coupling or the momenta at generic values.
A~distinctive feature of the exact pp-wave solution which was emphasized in \cite{SFTUS}
is that the large $L$ limit the Neumann coefficient $N_{33}(\th,\th')$ looses monodromy.
This has to be understood in the following sense. The exact function $n(\th)$ satisfies
\eq
n(\th+2\pi i) = e^{-ip(\th)L}\, n(\th)
\eqx
However if we first take the large $L$ asymptotics (c.f. (\ref{e.ntlas}), which is valid for $\Im m(\th) \in (0, 2\pi)$),
we get
\eq
n_{as}(\th+2\pi i) = - n_{as}(\th)
\eqx
This can be reformulated as the statement that
\eq
\lim_{\eps \to 0^+} \lim_{L \to \infty} \f{n(\th+2\pi i-i\eps)}{n(\th+i\eps)} = -1 
\eqx
while if we take the limit $\eps \to 0^+$ first, we get of course
\eq
\lim_{\eps \to 0^+} \f{n(\th+2\pi i-i\eps)}{n(\th+i\eps)} = e^{-ip(\th)L}
\eqx
for real $\th$.

We will now establish that a similar property holds for the $AdS$ kinematical Neumann coefficient.
Recall the structure of the corresponding quantity
\eq
n(z)=\f{2g \sqrt{L}}{\pi} \sin p\; G_L^{ev}(z) h^{ev}_L(z)
\eqx
Since $h^{ev}_L(z)$ is essentially just a $L^{th}$ power of a combination of the $L$-independent elliptic Gamma functions,
its asymptotic limit is trivial and its asymptotic monodromy coincides with the normal one i.e.
\eq
\label{e.hLratio}
\f{h^{ev}_L(z+\tau-\eps \tau)}{h^{ev}_L(z+\eps \tau)} \to e^{-ipL}
\eqx
for real $z$. We thus have to show that the corresponding ratio of $G_L^{ev}(z)$ has the opposite monodromy
\eq
\label{e.GLratio}
\f{G^{ev}_L(z+\tau-\eps \tau)}{G^{ev}_L(z+\eps \tau)} \to e^{ipL}
\eqx
when we first take $L$ large.

Let us focus on checking this condition when $z$ is real and belongs to the interval $z \in(0,1/2)$. Then
for points slightly above the real axis we have $E(z+\tilde{\eps}\tau) = E(z)+i\eps$ with both $\tilde{\eps}$ and
$\eps$ positive. Similarly we will have $E(z+\tau-\tilde{\eps}\tau) = E(z)-i\eps$. Thus the ratio (\ref{e.GLratio})
becomes
\eq
\prod_{k=1}^{n-1} \f{  \sqrt{1+16g^2 \sin^2 \f{\pi k}{L}}- E+i\eps }{  \sqrt{1+16g^2 \sin^2 \f{\pi k}{L}}- E-i\eps }
\eqx
and $L=2n$. Let us transform the product into an exponent of a sum of logarithms and take the large $L$ limit
by rewriting the discrete sum as an integral\footnote{This would be the leading term in a finite Poisson resummation.
We will not control subleading terms so we will not determine order 1 terms in the monodromy (like the $-1$ in the 
asymptotic pp-wave case).}. We get
\eq
\exp \left\{ \f{L}{2} \int_0^1  \log \left(\sqrt{1+16g^2 \sin^2 \f{\pi y}{2}}- E+i\eps \right) - \log\left(\sqrt{1+16g^2 \sin^2 \f{\pi y}{2}}- E-i\eps \right) dy
\right\}
\eqx
Now we have to be careful concerning the sign of the argument of the logarithm. When
\eq
y<y_* \equiv \f{2}{\pi} \arcsin \f{1}{4g} \sqrt{E^2-1}
\eqx
the argument is negative and we deal essentially with the discontinuity across the branch cut which is $2i \pi$.
Thus we get
\eq
\exp \left\{ \f{L}{2} \cdot 2i\pi \cdot y_* \right\}
\eqx
However using the dispersion relation $E=\sqrt{1+16 g^2 \sin^2 p/2}$ we see that $y_*=p/\pi$.
This gives finally
\eq
e^{i p L}
\eqx
which exactly cancels (\ref{e.hLratio}).

\section{Conclusions}

The String Field Theory (SFT) vertex axioms for $AdS_5 \times S^5$ are very challenging to solve for two reasons.
Firstly, they incorporate a nondiagonal S-matrix which even for the case
of ordinary relativistic form factors severly complicates their solution.
Secondly, they involve the dependence on the size of the third closed string $L$ which
leads to the incorporation of all order multiple wrapping effects w.r.t. this parameter.
This is in contrast to the hexagon approach where the question of wrapping is dealt with
iteratively on top of an exact asymptotic solution. On the other hand, the possibility of handling
analytically an infinite set of wrapping corrections at once is very appealing.

In this paper we have analyzed the two-particle SFT vertex axioms and found that
one can factor out the complicated dependence on the dynamical S-matrix into
an $L$-independent form factor\footnote{Its determination still remains as an outstanding open problem. See 
\cite{KM2,KM1} for details.}
and a piece that satisfies the $L$\emph{-dependent} axioms but with $S=1$.
We refer to this solution as the `kinematical Neumann coefficient'.
By definition this solution includes an infinite set of wrapping corrections
w.r.t. $L$ which are necessary to solve the SFT vertex axioms.
Of course, one can conceive adding some additional $L$ dependence in a form
of an analog of a CDD factor, but since $L$ does not appear in the CDD equations
this is not enforced by any equations. So in this sense the `kinematical Neumann coefficient'
is a minimal solution as far as wrapping is concerned.

In this paper we have constructed explicitly the kinematical Neumann
coefficient in the case of $AdS_5 \times S^5$ kinematics for any even $L$
and any value of the gauge theory coupling constant.
We have verified that this expression has the correct pp-wave limit and that
it reduces to the spin chain answer (up to an $L$-independent CDD factor).
In addition we have analyzed the large $L$ limit and verified that
it obeys analogous properties to the known pp-wave solution
namely the apparent cancelation of monodromy in the physical strip.

It would be very interesting to solve the form factor equations in $AdS_5 \times S^5$ in order to
obtain a complete expression. Also it is important to understand whether any CDD factors
are necessary for a physical solution. Especially whether any additional wrapping effects would
have to be included in the CDD factors.
We hope that this question could also be addressed in
the simpler relativistic setting where there exist integrable lattice realizations of 
some theories. To facilitate that we quote the complete solutions of the two-particle SFT vertex axioms
for the sinh-Gordon and O(N) model.

The weak coupling analogue of the string vertex is the spin vertex. In  \cite{Jiang:2014cya,Kazama:2014sxa} the 
authors constructed the finite volume (size) spin vertex for all sectors at leading order. It
would be interesting to investigate the decompactification limit of the vertex, which corresponds 
to our geometry and see how the vertex factorizes into a kinematical and a form factor part. 
This leading order calculation could be extended to higher loops by relating the higher order 
long-ranged spin chains to inhomogenous spin chains \cite{Jiang:2014mja} or by using the 
separation of variables basis \cite{Jiang:2015lda}. However, to directly compare the spin vertex
of \cite{Jiang:2014cya} with the string vertex one would have to include in the former the contribution
of Bethe wave functions of the external states. A comparision has so far been done only in the pp-wave
limit \cite{Jiang:2014gha}.

Finally we hope that the exact treatment of multiple wrapping corrections
in the kinematical Neumann coefficient formula will be helpful for the treatment of wrapping in
the hexagon approach.

\bigskip

{\bf Acknowledgements.} 
RJ was supported by NCN grant 2012/06/A/ST2/00396 and ZB by a Lend\"ulet Grant. 
This work was supported by a Polish-Hungarian Academy of Science
cooperation.

\end{document}